\documentclass[aps,pre,twocolumn,amsmath,amssymb,10pt]{revtex4-1}

\usepackage{amssymb,amsfonts,amsmath,graphicx}
\usepackage{color}
\usepackage{natbib}
\usepackage{float}
\usepackage{bm}

\def\be{\begin{equation}}
\def\ee{\end{equation}}
\def\bea{\begin{eqnarray}}
\def\eea{\end{eqnarray}}

\def\l{\left(}
\def\r{\right)}

\def\be{\begin{equation}}
\def\ee{\end{equation}}
\def\bea{\begin{eqnarray}}
\def\eea{\end{eqnarray}}

\def\ra{\rangle}
\def\la{\langle}

\def\x{{\bf x}}

\def\V{{\bf V}}

\def\la{\langle}
\def\ra{\rangle}
\def\l{\left(}
\def\r{\right)}

\def\l{\left(}
\def\r{\right)}

\usepackage[
margin=0.7in,
]{geometry}
\begin{document}

\title{Interacting passive advective scalars in an active medium
}

\author{SK Raj Hossein}
\affiliation{%
Raman Research Institute, Bangalore 560080, India
}%
\author{Rituparno Mandal}
\affiliation{%
Centre for Condensed Matter Theory, Department of Physics,
 Indian Institute of Science, Bangalore 560012, India
}%
\author{Madan Rao}
\email{madan@ncbs.res.in}
\altaffiliation{%
on lien from : Raman Research Institute, C.V. Raman Avenue, Bangalore 560080, India
}%
\affiliation{%
Simons Centre for the Study of Living Machines, National Centre for Biological Sciences (TIFR), Bellary Road, Bangalore 560065, India
}%

\begin{abstract}
Recent experimental studies, both {\it in vivo} and {\it in vitro}, have revealed that membrane components that bind to the cortical actomyosin meshwork are driven by active fluctuations, whereas 
membrane components that do not bind to cortical actin are not.
 Here 
we study the statistics of density fluctuations and dynamics of particles advected in an active {\it quasi}-two dimensional medium 
comprising self-propelled filaments with no net orientational order, using a combination of agent-based Brownian dynamics simulations and analytical calculations. The particles interact with each other and with the
self-propelled active filaments via steric interactions. 
We find that the particles show a tendency to cluster and their density fluctuations reflect their binding to and driving by the active filaments.
The late-time dynamics of tagged particles 
is diffusive, with an {\it active diffusion} coefficient that is independent of (or at most weakly-dependent on) temperature 
at low temperatures. Our results are in qualitative agreement with 
the experiments mentioned above. In addition, we make predictions that can be tested in future experiments.
\end{abstract}


\maketitle

\section{Introduction and Motivation}

The spatial organization, clustering and dynamics of many cell surface molecules is influenced by interaction with 
the actomyosin cortex~\cite{raomayor,Goswami,kripa,saha,Hancock,Lingwood}, a thin layer of actin cytoskeleton and myosin motors, measured to be around $250$\,nm in HeLa cells~\cite{Paluch}, that is 
juxtaposed with the membrane bilayer. Although the ultrastructure of the cortical actin cytoskeleton is as yet poorly defined, there is growing
evidence that it is composed simultaneously of dynamic filaments~\cite{kripa} and an extensively branched static  meshwork~\cite{Morone}. 
The coupling of the membrane to these two types of actin configurations is expected to affect the dynamics and
organization of membrane components, presumably in different ways.

\begin{figure}
\hspace{2mm}
\includegraphics[height=0.55\linewidth]{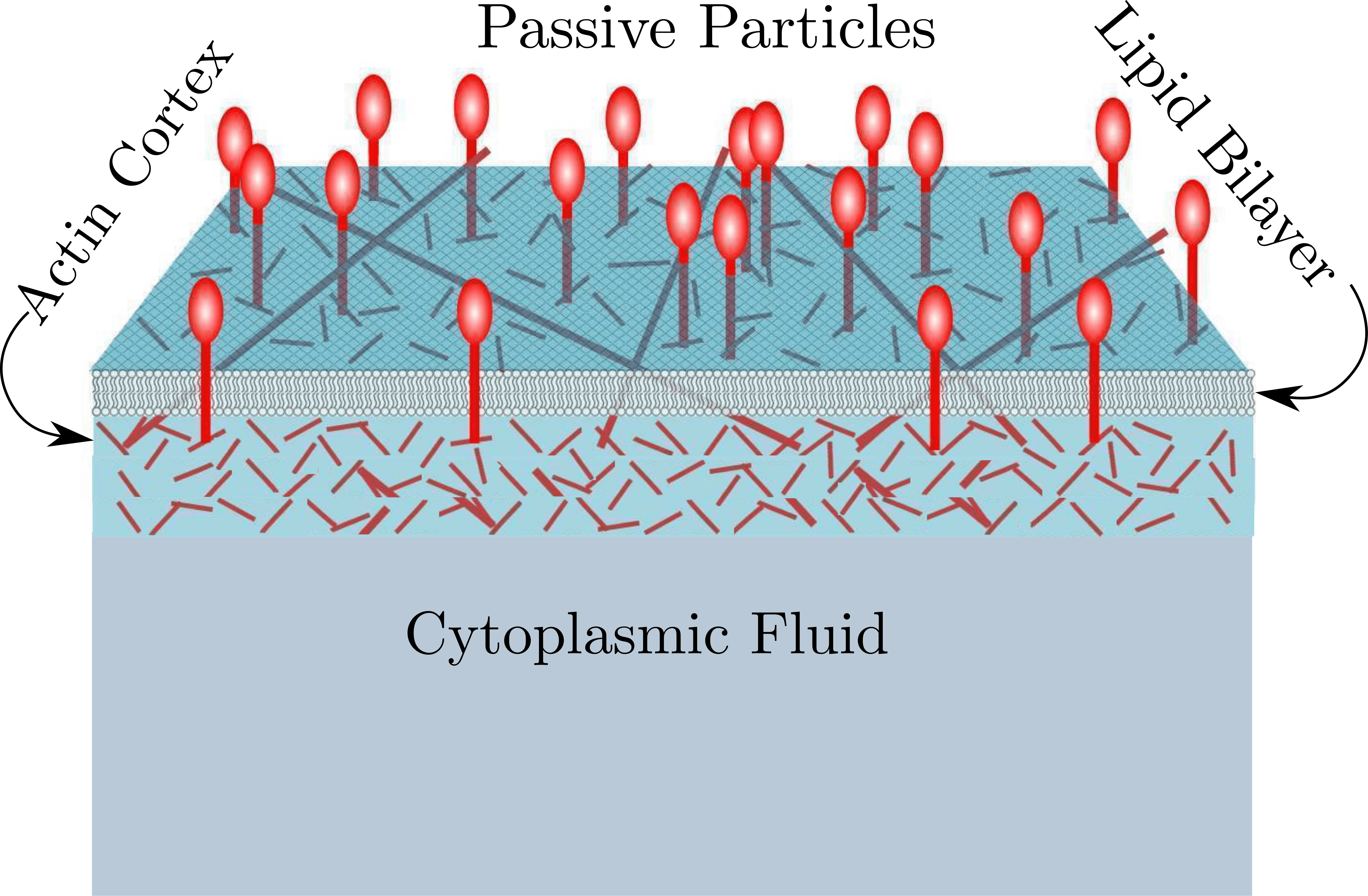}
\caption{Schematic of the Active Composite Cell Surface with a multicomponent asymmetric bilayer
membrane juxtaposed with the cortical layer of actin and myosin. The actin layer consists of 
short dynamic actin and long static actin filaments, the latter forming a crosslinked mesh~\cite{Morone,fujiwara}.
For clarity, myosin motors are not shown, the short filaments are active and represent myosin bound actin filaments. The passive molecules (shown in red)
are transmembrane proteins that have a cytoplasmic actin-binding domain.}
\label{fig:schema1}
\end{figure}

A common feature of these cell surface proteins is that they can bind, directly or indirectly, 
to cortical actin. As a consequence, the action of myosin motors on actin at the cortex help drive the local clustering and dynamics of these cell surface proteins.
Mutations of these proteins that abrogate this actin binding capacity, leave them unaffected 
by the dynamics of the actomyosin cortex~\cite{kripa,saha}. 
Similarly, silencing myosin motor activity, renders the dynamics of these cell surface proteins normal and akin to their mutated counterpart~\cite{kripa,saha}.
 A description of the cell surface as an {\it Active Composite} of
a multicomponent, asymmetric bilayer juxtaposed with a thin cortical actomyosin layer (Fig.\,\ref{fig:schema1}), appears to
consistently explain the anomalous dynamical features of these proteins~\cite{raomayor,kripa,saha}. In these earlier studies 
~\cite{kripa,kripasoft}, we had used a coarse grained description, based on active hydrodynamics~\cite{rmp}.
Recently, it has been shown that much of this behaviour is recapitulated in a minimal 
{\it in vitro} reconstitution of a supported bilayer in contact with a thin layer of short actin filaments 
and Myosin-II minifilaments, driven by the hydrolysis of ATP~\cite{darius}. The success of this minimal setup motivates us 
to revisit our continuum hydrodynamic description from a more microscopic standpoint.
Our present study is an agent-based Brownian dynamics simulation of a mixture of
polar active filaments and passive particles which interact with each other. Since the parameter space for exploration is large, 
we restrict our study here to the case where the density of filaments and particles is low (negligible filament overlap), 
and where there is no net orientational order.

For the present purposes, we schematically represent the Active Composite Cell Surface, as in Fig.\,\ref{fig:schema1}.
To enable a systematic study, we conceptually separate out the different architectures of actin at the cortex -
(i) where the cortical layer consists of only short dynamic filaments described as an active fluid (the subject of the 
present paper) and (ii) where the cortical  layer consists of only long filaments forming a static mesh of 
characteristic mesh size $\xi \approx 52$\,nm in FRSK cells~\cite{Morone,fujiwara} (which we take up in a later publication). 
In future, we will combine these configurations into a single model of the cortical actomyosin.

 Our choice of Brownian dynamics simulations
  is motivated by experiments on tagged particle diffusion both on the cell surface and in the {\it in vitro} reconstitution.
 Molecules that bind to dynamic actin (passive molecules) are affected by the 
active fluctuations of actomyosin - their diffusion shows anomalous behaviour strongly indicative of active driving. On the other hand, 
molecules that do not interact with actin (inert molecules), such as short chain lipids and proteins whose actin-binding domain has been 
mutated so as to abrogate their interaction with actin, do not show any influence  of active fluctuations~\cite{kripa,saha,darius}. There appears to be no sign that the transport of these
inert molecules is affected by potential 
hydrodynamic flows induced by active stresses coming from actomyosin~\cite{yhat,abasu}.

While our primary motivation are the experimental studies of the tagged particle dynamics on the cell surface, our work is also relevant to transport in other
living and nonliving systems, as long as the effects of hydrodynamics are negligible, for instance, to the movement of multiple motor-driven cargo vesicles or synthetic beads on the cytoskeletal 
network~\cite{Grannick}.

\section{Brownian dynamics and characterization}

\subsection{Simulation details}

We study the dynamics of a mixture of polar active filaments and passive particles using 
an agent-based Brownian dynamics simulation. 
The passive particles are modelled as mono-disperse soft spheres of diameter $\sigma$. A pair of passive particles separated by a distance $r$ interact via a truncated Lennard Jones (LJ) pair potential
of the form,
\bea
V_{pp}(r) & = & 4 \epsilon \Bigg[\bigg(\frac{\sigma}{r}\bigg)^{12}-\bigg(\frac{\sigma}{r}\bigg)^{6}\Bigg] + V_0 +V_2 r^2  \,\,\,\,\, \mbox{for} \,\, r\leq r_c  \nonumber \\
        & =  &   0    \,\,\,\,\, \mbox{for} \,\, r > r_c 
        \label{partpot}
\eea
where $r_c=2.5 \sigma$ and values of $V_0$ and $V_2$ are chosen so that the potential and force are continuous at the truncation point. 
We set $\sigma=1$ and $\epsilon=1$ to be the units of length and energy, respectively.

The polar filaments are modelled as semi-flexible bead-spring polymers, with 
both stretch and bend distortions.
We implement excluded volume interaction between the beads of same filament,
as well as between two different filaments through a truncated Lennard Jones pair potential of the form,
\bea
V_{bb}(r) & = & 4 \epsilon^{\prime} \Bigg[\bigg(\frac{\sigma^{\prime}}{r}\bigg)^{12}-\bigg(\frac{\sigma^{\prime}}{r}\bigg)^{6}\Bigg] +V_{0}^{\prime}+ V_{2}^{\prime} r^2\,\,\,\,\, \mbox{for} \,\, r\leq r^{\prime}_c  \nonumber \\
       & = & 0   \,\,\,\,\, \mbox{for} \,\, r > r^{\prime}_c
           \label{beadpot}
\eea
where 
$r$ is the distance between the centres of the corresponding beads, 
$r^{\prime}_c=2^{1/6} \sigma^{\prime}$ and $V_{0}^{\prime}, V_{2}^{\prime}$ are constants, 
chosen so that the potential and force are continuous at $r^{\prime}_c$.
We take $\sigma^{\prime}=2$ and $\epsilon^{\prime}=1$. Each filament is composed of
$10$ beads and therefore has an equilibrium length $l_0=20$, in the units of $\sigma$.

Note that with our choice of cutoffs, the particle-particle interaction $V_{pp}$ has both attractive and repulsive parts, whereas the 
bead-bead interaction $V_{bb}$ is strictly repulsive.

To make the filament semi-flexible, we impose additional spring forces on the beads. 
A harmonic stretching potential with 
spring constant $K_c=400$, in units of $\epsilon/\sigma^2$, ensures that the length of the filament does not deviate significantly from its
equilibrium value, $l_0=20$. The bending energy of a triplet
of connected beads is also harmonic in the angle, with a bending stiffness 
$K_a = 600$, in units of $\epsilon$. This high $K_a$ makes the filaments very stiff, with 
a typical persistence length much larger than $l_0$.

A propulsion force ${\bf f}=f_0 {\bf {\hat n}}$ is imposed 
on each of the beads, along the average direction (${\bf {\hat n}}$) of all the bonds present in a filament. 
Note that we do not impose any filament alignment rule nor do we prescribe any activity decorrelation time. 
Instead, these originate from thermal fluctuations on the constituent monomers comprising each filament and 
collisions driven by thermal and active forces, an emergent many-particle feature. As a consequence, 
both the local alignment and orientational de-correlation time are functions of 
temperature, density  and activity.

The interactions between the beads of the filament and the passive particles are modelled by a 
harmonic potential of spring constant $K_s=50$ in units of $\epsilon/\sigma^2$. The harmonic potential is
truncated at a cutoff distance $r_{0}= 1$ and set to zero beyond it. When a passive
particle comes within a distance $r_{0}$ from the centre of a filament bead, it binds 
to the corresponding bead and gets advected along with the filament, under the application of propulsion force $f_0 {\bf {\hat n}}$.

Unless mentioned otherwise, all results presented here are for $N_p=800$ passive particles 
and $N_r=50$ self-propelled filaments in a two dimensional area of linear dimension $L=396.4$. For most of the study, we
take the area fractions of the rods ($c$) and particles ($\rho$) to be $c = 0.01$ and $\rho=0.004$, respectively.

The Brownian dynamics equations involve an update of both the passive particle and the filament bead coordinates, for which we have used a simple Euler integration scheme with
integration time step $\Delta t \sim 10^{-4}$.
The dynamics of the particle coordinates is given by

\be
    \dot {\bf r}_i^p=
    \begin{cases}
      -\gamma^{-1}_p {\boldsymbol \nabla}_i V_p   +  \sqrt{2 k_BT/\gamma_p}{\boldsymbol \xi}_i, & \text{(unbound)}\ \\
      -\gamma^{-1}_p {\boldsymbol \nabla}_i V_p   +  \sqrt{2 k_BT/\gamma_p}{\boldsymbol \xi}_i+ {\bf f}_i/\gamma_p, & \text{(bound)}
    \end{cases}
    \label{particle}
\ee
where
$\gamma_p$ is the friction coefficient of the passive particle, $V_p$ is the net potential felt by the $i$-th passive particle and includes contributions 
from Eq.\,\ref{partpot} and the bead-particle spring interactions.
The diffusion of the unbound particle is driven by a thermal noise
${\boldsymbol \xi}_i$  with zero mean and unit variance acting on $i$-th particle.
On the other hand, the bound particles are subject to both the thermal noise and active driving.
 
The dynamics of the filament-bead coordinates is given by
\be
 \dot {\bf r}_j^b = -\gamma^{-1}_b {\boldsymbol \nabla}_j V_b   + \sqrt{2 k_BT/\gamma_b} \,{\boldsymbol \xi}_j +  {\bf f}_j/\gamma_b                                        \,
\label{bead}
\ee
where
$\gamma_b$ is the friction coefficient of the bead, $V_b$ is the net potential felt by the $j$-th bead and includes contributions from Eq.\,\ref{beadpot}, harmonic stretching and bending interactions.

 We take $\gamma_p=1$, which together with $\sigma=1$ and $\epsilon=1$, sets the units of space, time and energy. 
 All other quantities can be written in terms of these units, so as to make Eqs.\,\ref{particle}, \ref{bead} dimensionless.
In all that follows below,  except in Sec.\,V,  we have taken $\gamma_b=\gamma_p$.
 To be able to make contact with experiments, we have to translate our simulation units (S.U.) to
real units (R.U.). Setting $\sigma = 10$\,nm,  $\gamma=0.123$\,pN ${\mu m}^{-1}$ s~\cite{kamm} and $\epsilon = 4.14 \times 10^{-21}$\,J, 
we can convert our simulation units to real units, as displayed in Table\,1.
 
We have typically run the Brownian dynamics simulation for a total time $t \sim 10^4$, ensuring that 
the system has reached steady state.
We have varied the temperature over $T = 0.25, 0.5, 0.75, 1, 2, 4, 6, 8, 10, 15, 25$ and the
propulsion force over $f_{0} = 0.0, 0.5, 1.0, 2.0, 3.0, 4.0$. Our initial conditions are chosen
from a thermal distribution at temperature $T$ and all results presented here are averaged over $16$ such independent
initial realisations.

\subsection{Statistics of filament orientation}
\begin{figure}
\includegraphics[height=0.37\linewidth]{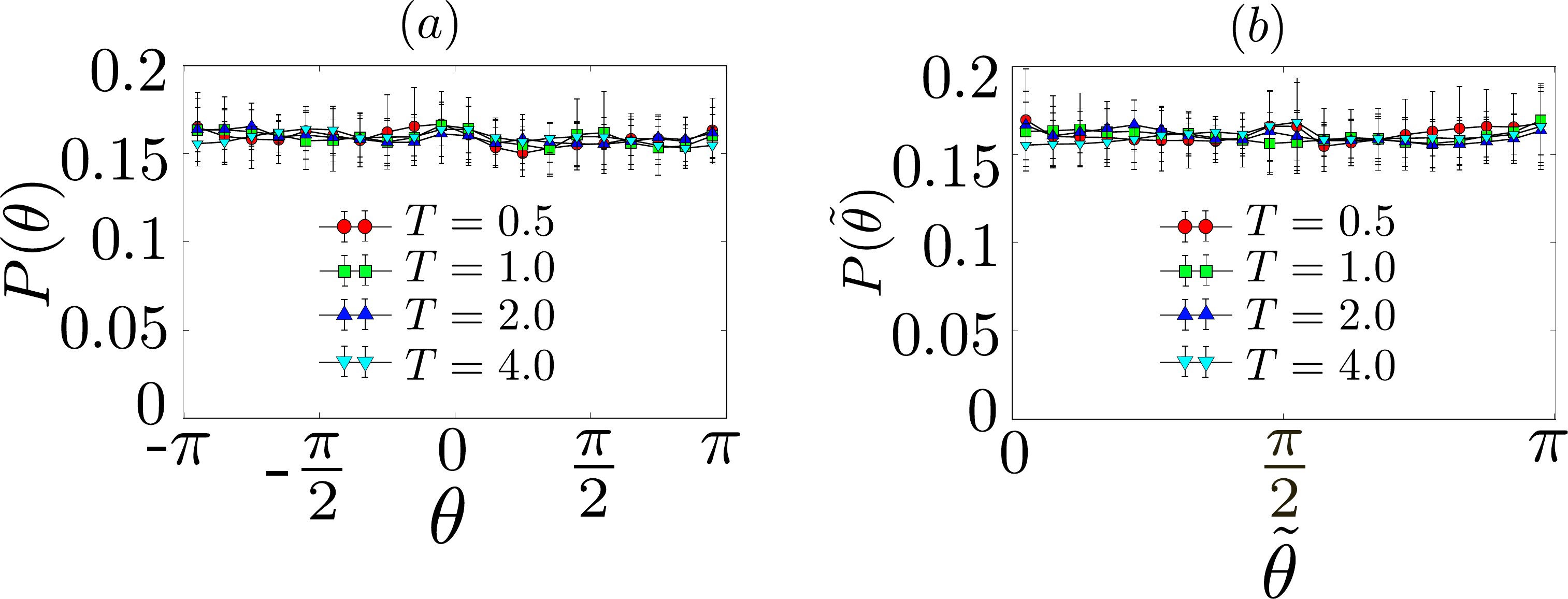}
\caption{Normalised distribution of (a) polar orientation $P(\theta)$ and
(b) nematic orientation $P({\tilde \theta})$, of the filaments at different temperature $T$ with activity $f_0=4.0$, showing that the system is in the isotropic phase for a representative set of parameters.} 
\label{theta_dis}
\end{figure}

We characterise the $i$-th filament by its  centre of mass position ${\bf r}_i $ and a unit vector ${\bf n}_i =(\cos\theta_i , \sin\theta_i )$ 
along its long axis to describe its polar orientation (recall that the filaments are very stiff). 
We first ensure that the configuration of filaments is in the spatially homogeneous, orientationally isotropic state - this is demonstrated in 
the plots of the probability distribution of the polar $P(\theta)$ and nematic orientations $P(\tilde \theta)$ (Fig.\,\ref{theta_dis}).

\begin {table}
\caption {Conversion between simulation units (S.U.) and real units (R.U.)}
\begin{center}
 \begin{tabular}{||c ||c ||c ||} 
 
 \hline
 $Parameter [Dimension]$       & $ S.U. $ &   $R.U.$                      \\ [0.5ex] 
 \hline\hline
 
  $\sigma$ [$l$]  & 1 &   $ 10$                 $nm$      \\ 
 \hline 
  
  $\epsilon$ [$m l^2 t^{-2}$]  & 1 &  $4.14 \times 10^{-21}$           $J$     \\ 
 \hline
  
  $\gamma_p$ [$m t^{-1}$]  &  1  &   $0.123$        $pN {\mu m}^{-1} s$        \\ 
 \hline
 
 $T$ [$k$]  & 1 &   $300$   $K$        \\ 
 \hline

  $t$ [$t$]   &  1  &   $3 \times 10^{-3}$           $s$        \\ 
 \hline
 
  $k_u,k_b$ [$t^{-1}$]   & 1  &   $333$      $s^{-1}$        \\ 
 \hline
 
 $f_0$ [$mlt^{-2}$]    & 1  &   $0.41$ $pN$        \\ 
 \hline

 $V_a$ [$lt^{-1}$]    & 1  &   $3.3 $ $\mu m$ $ s^{-1}$        \\ 
 \hline
 
  $K_c$ [$mt^{-2}$]   &  1  &   $41.4 $ $pN {\mu m}^{-1}$        \\ 
 \hline
 
  $K_a$ [$ml^2t^{-2}$]   &  1  &   $4.14 \times 10^{-21}$ $J$        \\ 
 \hline
 
  $K_s$ [$mt^{-2}$]   &  1  &   $41.4 $ $pN {\mu m}^{-1}$        \\ 
 \hline
 
 $D$ [$l^2t^{-1}$]             &   1   &   $3.3 \times 10^{-2}$                 ${\mu m}^2/s$ \\ 
 \hline 
\end{tabular}
\end{center}
\end{table}

\begin{figure}
\includegraphics[height=0.62\linewidth]{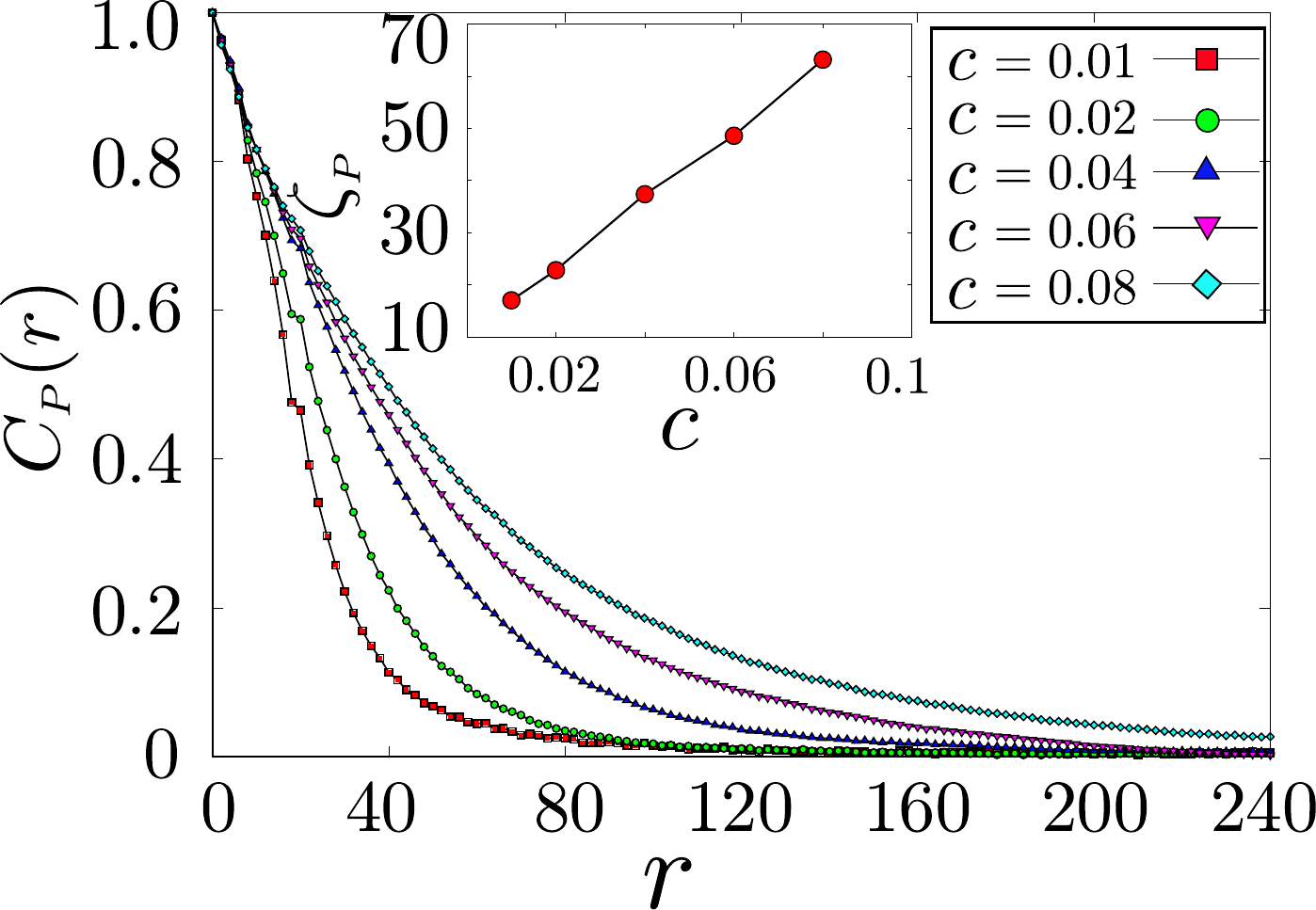}
\caption{Spatial correlation of the polar orientation, $C_P(r)$, of the filaments at different filament densities $c$ 
 at $T=0.5$ and active propulsion $f_0=4$.  Inset shows the corresponding correlation length $\zeta_P$ as a function of filament density $c$.}
\label{polarcorr}
\end{figure}

\begin{figure}
\includegraphics[height=0.65\linewidth]{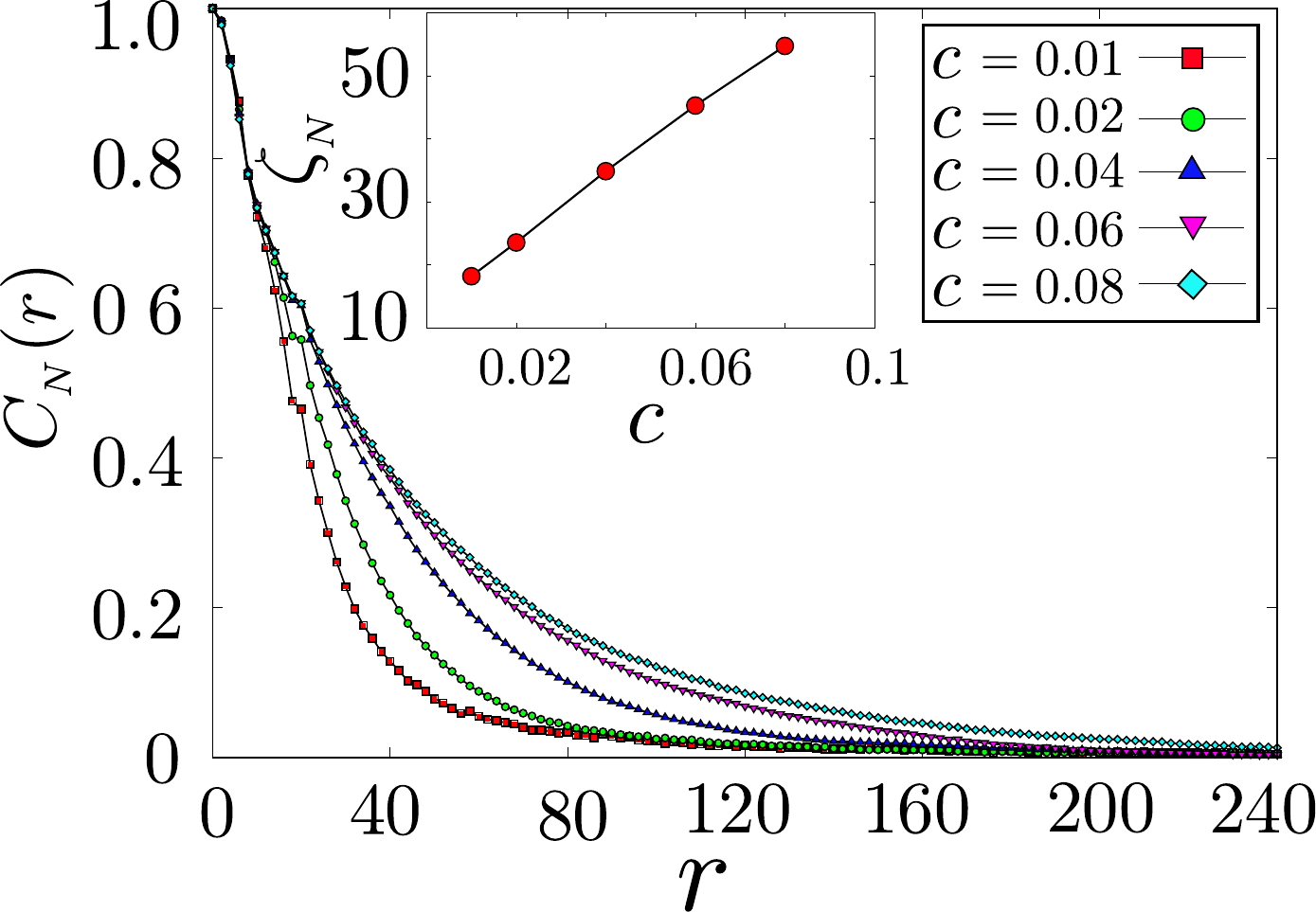}
\caption{Spatial correlation of the nematic orientation, $C_N(r)$,  of the filaments at different filament density $c$ at $T=0.5$ and 
$f_0=4$.  Inset shows the corresponding correlation length $\zeta_N$ as a function of filament density $c$. }
\label{nematiccorr}
\end{figure}

We then calculate the orientational correlation lengths, so as to ensure that this is much smaller than our system size and comparable to the size of the filaments. To do this, we
 calculate the spatial correlations  of both the polar and nematic orientation,
\bea
C_P(r) & = & \Bigg\langle \frac{1}{N^2} \sum_{i=1}^N \sum_{j=1}^N  \cos(\theta_i -\theta_j) \Bigg\rangle \\
C_N(r) & = &\Bigg\langle  \frac{1}{N^2} \sum_{i=1}^N \sum_{j=1}^N ( 2  \cos^2(\theta_i -\theta_j) -1 ) \Bigg\rangle
\eea
where $r= \vert {\bf r}_i - {\bf r}_j \vert$ is the distance between the centre-of-mass of the $i$-th and $j$-th filaments.
By fitting this to an exponential (Figs.\,\ref{polarcorr}\,and\,\ref{nematiccorr}), we extract the polar and nematic orientation correlation lengths, $\zeta_P$ and $\zeta_N$,
whose dependence on the area fraction of  filaments  $c$ (we will henceforth refer to this as filament density) is shown in the inset.

Throughout the paper (unless mentioned otherwise) we work at a filament density of $c=0.01$, and temperatures $T\geq0.5$; in this regime, the
orientational correlation lengths are of the order of the filament length, $l_0$, and hence comfortably within the isotropic phase.

\subsection{Statistics of filaments persistence}
\begin{figure}
\includegraphics[height=0.65\linewidth]{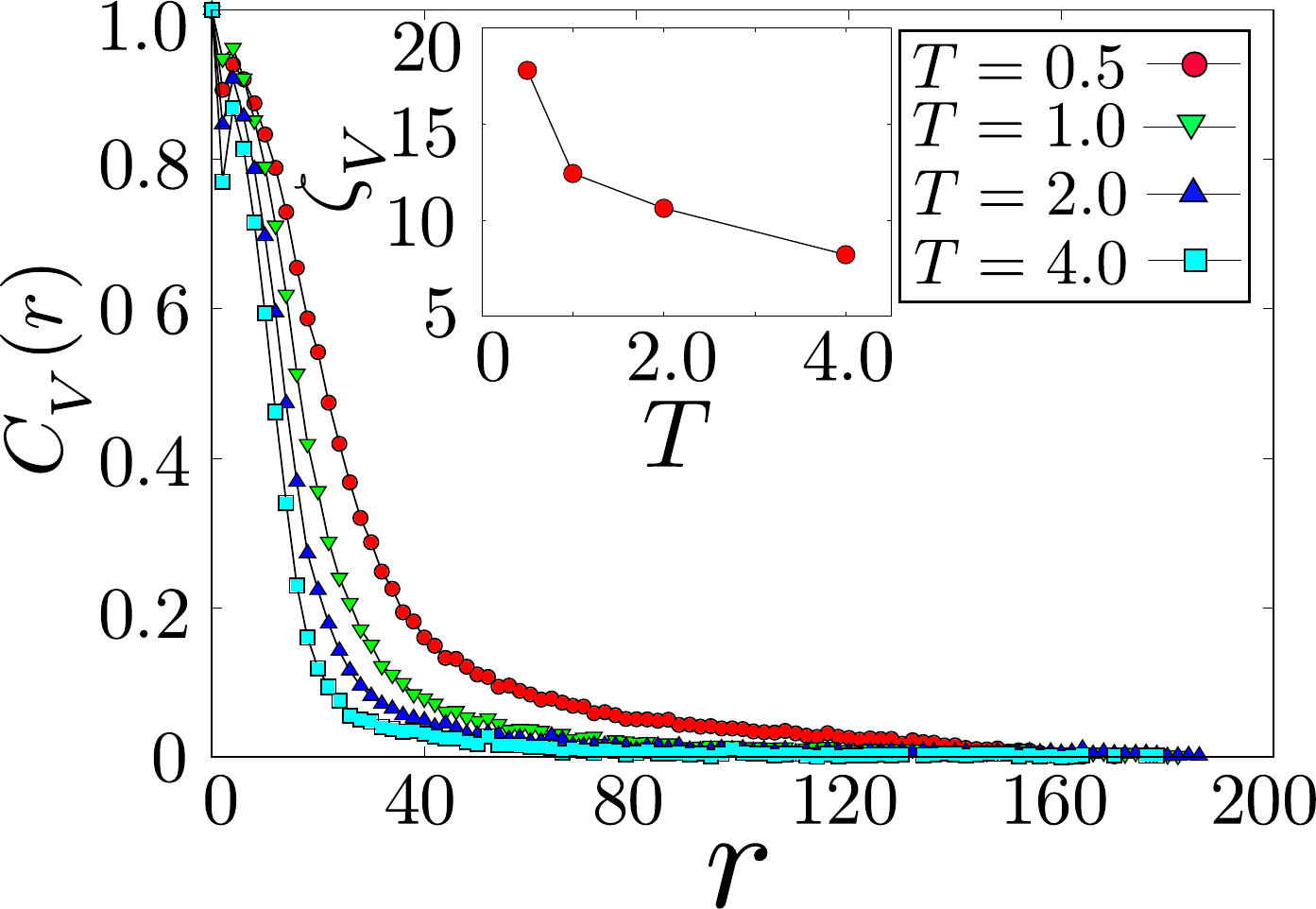}
\caption{Spatial correlation of the direction of persistent movement of the filaments, measured by the velocity orientation correlation function $C_V(r)$, at different temperatures, with
the filament density $c =0.01$ and propulsion force $f_0=4$.  Inset shows the corresponding correlation length $\zeta_V$ as a function of $T$. }
\label{velcorr}
\end{figure}

A self-propelled filament moves persistently in a direction set by ${\bf V}_a=\frac{f_0 \bf{\hat{n}}}{\gamma_b}$, taken to be along its polar orientation, until its orientation gets decorrelated, either due to thermal noise on the individual monomers constituting the
filaments or due to inter-filament 
collisions. What is the spatial correlation of these directions of persistent motion? Here, we calculate this velocity orientation correlation function $C_V(r)$ at different $T$ (Fig.\,\ref{velcorr}) using the formula
\bea
C_V(r) & = &\Bigg\langle  \frac{1}{N^2} \sum_{i=1}^N \sum_{j=1}^N ( 2  \cos^2(\Theta_i -\Theta_j) -1 ) \Bigg\rangle
\eea
where $\Theta_i$ represents the direction of the velocity vector of the $i$-th filament.
From this spatial correlation function $C_V(r)$ we extract a 
correlation length ($\zeta_V$) which measures the spatial extent of the persistent dynamics. We find again that at $c=0.01$ over the temperature range $T \geq 0.5$,
 the correlation length $\zeta_V$ is comparable with the filament length $l_0$.
 
\subsection{Statistics of binding-unbinding of passive particles onto filaments}

The dynamical equations (\ref{particle}) and (\ref{bead}) are written entirely in terms forces, either active or derived from a potential, and thermal noise. The particles experience a 
binding and unbinding onto the filaments which depend on this interplay between thermal noise and the attractive potentials. Thus for instance, 
the unbound passive particles diffuse in the two dimensional medium and ever so often come within the vicinity ($r \leq r_0 = 1$) of a moving filament-bead, whereupon they bind to the filament-bead.
In the low density limit, we expect the binding rate $k_b$ to be diffusion limited and so $k_b \propto T$ and independent of $K_s$ where $K_s$ is the strength of trapping harmonic interaction.

To study the unbinding of a particle bound from a filament-bead, we compute the rate of escape of a particle trapped in a truncated attractive harmonic potential~\cite{grebenkov},
parametrised by $K_s$ and $r_0$. This is given by
\be
k_u = \frac{K_s^2 r_0^2}{\gamma_p k_B T} \, \exp\left(-  \frac{K_s r_0^2}{2 k_B T}\right)
\label{ku}
\ee
and should be a good description of the dynamics of unbinding of the passive particles in the limit of low particle density.

We compare these theoretical estimates with the results of simulations
on a mixture of particles and filaments at equilibrium (no active propulsion), from which
we extract the values of $k_u$ and $k_b$ in two different ways. In the first method,
we represent the stochastic binding and unbinding by a 
telegraphic process~\cite{kampen}, characterised by a mean duty ratio,
\be
 \la \phi \ra = \frac{k_b }{k_b + k_u}\, ,
\ee
the fraction of
time spent by the tagged particle in the  bound state over the observation time, and a 
 two-point correlator,
\be
\la  \phi(t) \phi(t^{\prime}) \ra =  \la \phi \ra^2 +  \la \phi \ra \l 1- \la \phi \ra \r e^{-2 \frac{\vert t-t^{\prime}\vert}{t_{sw}}}
\label{2pt}
\ee
where 
\be t_{sw} = \frac{2}{k_b+k_u} \, ,
\label{switch}
\ee
is called the mean switching time and describes the mean time taken to switch from a bound to an unbound state.
We  calculate $k_b$ and $k_u$, by fitting our simulation results to  $\la \phi \ra$  and $\la  \phi(t) \phi(t^{\prime}) \ra$. In the second method,
we calculate $k_u$ ($k_b$) directly, from the inverse mean time that the particle stays bound (unbound) on the filament.

\begin{figure}
\begin{center}
\includegraphics[height=0.36\linewidth]{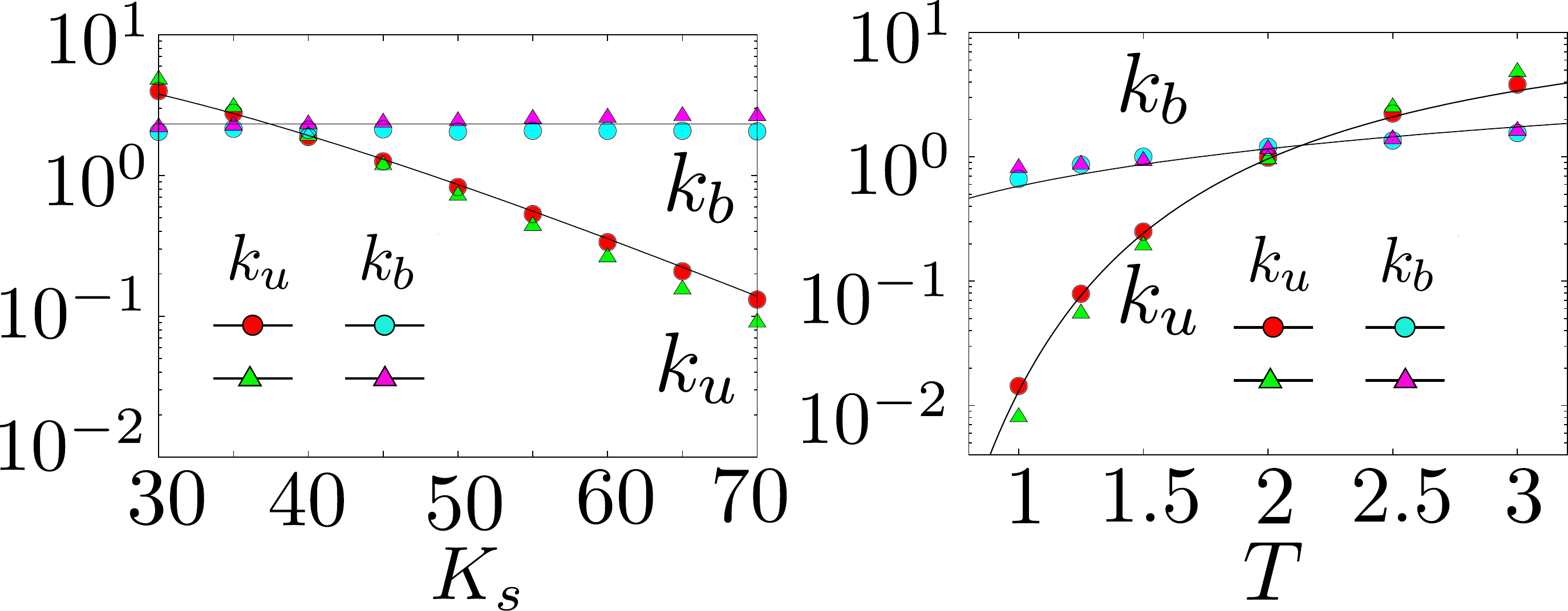}
\caption{Dependence of binding (unbinding) rates $k_b$ ($k_u$) on (a) the strength of the particle-filament binding potential, $K_s$,  and (b) temperature, $T$,
 calculated using the two  different numerical methods (filled symbols) discussed in the text. These can be fit, with no undetermined parameter,
 to the analytic forms (solid lines) discussed in the text (Eq.\,\ref{ku}).
}
\label{rate}
\end{center}
\end{figure}

The two numerical methods give identical results (after scaling by a constant factor) as a function of the particle-filament  binding potential $K_s$ and $T$, and that these
agree with the analytical estimates, Fig.\,\ref{rate}, with no fit parameter.

It is important that we do not {\it prescribe} the binding-unbinding rates, rather 
we derive 
them from the assigned potentials. The binding-unbinding rates thus depend nontrivially on temperature;  they would also depend on the
density of passive particles and filaments in the high density limit.
This will  be crucial to our estimation of the tagged particle diffusion coefficient and its comparison with experimental data (Sect.\,III\,D).

\section{Density fluctuations of passive advective scalars in an active medium}

We now study the statistics of density fluctuations and dynamics
of the actively driven passive particles. We find that the active driving tends to cluster the passive particles; this shows up in
 the two point spatial density correlation function and the statistics of the density fluctuations. 

\subsection{Radial distribution function}

\begin{figure}
\includegraphics[height=0.65\linewidth]{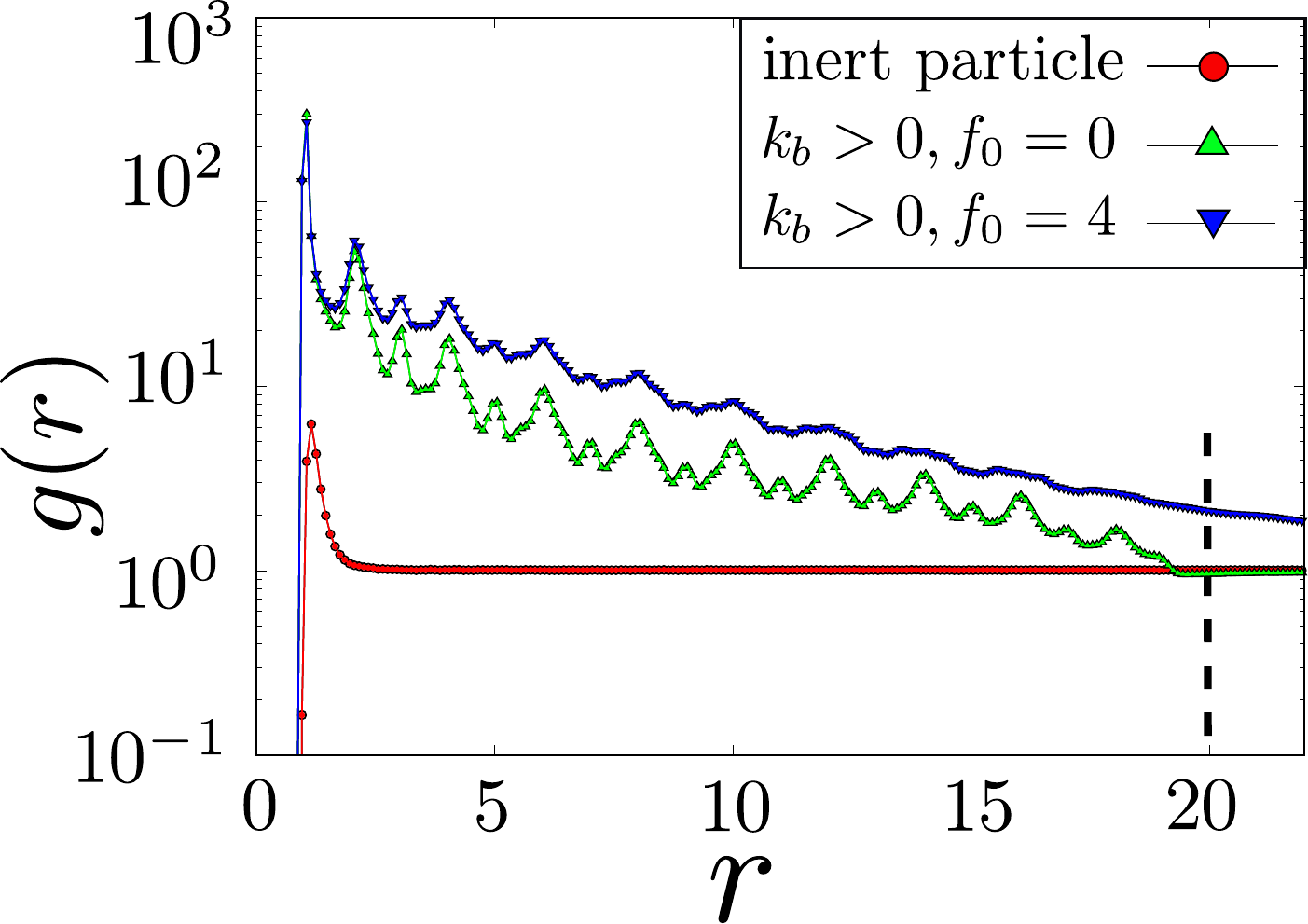}
\caption{Radial distribution function $g(r)$ for inert particles ($k_b=0$ : red $\bullet$), passive particles which can bind to the filaments at equilibrium ($k_b>0$, $f_0=0$ : green $\bigtriangleup$) and passive particles which can bind to the self-propelled filaments ($k_b>0$, $f_0=4$ : blue $\bigtriangledown$). In the presence of activity the peak heights increase and $g(r)$ falls off more gradually, indicating a high
degree of clustering in both the bound and unbound particles.
}
\label{gr_density}
\end{figure}

We study the behaviour of the radial distribution function of the passive particles $g(r)$,
\begin{equation}
g(r)= \frac {1}{N_p\rho}\Bigg\langle \sum \limits^{}_{i}\sum \limits^{}_{i\neq j} \delta(r-{\lvert \bf r_i - r_j \rvert}) \Bigg\rangle,
\end{equation}
where $N_p$ is the total number of passive particles and $\rho$  the passive particle density. 
When $k_b=0$ and $f_0=0$, i.e., when the particles do not bind to the filament (inert particles) and there is no propulsion force, $g(r)$ has the form of a dilute fluid (Fig.\,\ref {gr_density}).
When we allow for particle binding, but in the absence of propulsion force, the $g(r)$ displays 
oscillations, which arise from particles binding to periodic locations on the filaments (Fig.\,\ref {gr_density}) - note $r=20$ coincides with the filament  length. In this equilibrium situation, the 
 particles not bound to the filaments do not show any clustering.

We now consider the case when the filaments are driven by a propulsion force $f_0$. We find that in addition to the periodically spaced bound particles, there is a significant fraction of unbound particles that are clustered
 (Fig.\,\ref {gr_density}).
This clustering is a consequence of the shepherding of  passive particles by the active filaments - this has also been reported in~\cite{nitin}, with the difference that in that case the dynamics of
the particles affects the active filaments.

\subsection{Probability distribution of local number density} 
This activity induced clustering of the passive particles should be reflected in the probability distribution of the excess number density.
 To  compute this we
divide the  system into blocks of size $\Omega = 39.64$ and
count the number of passive particles $n$ in each block, to obtain the 
 steady state distribution $P(n)$.

\begin{figure}
\includegraphics[height=0.68\linewidth]{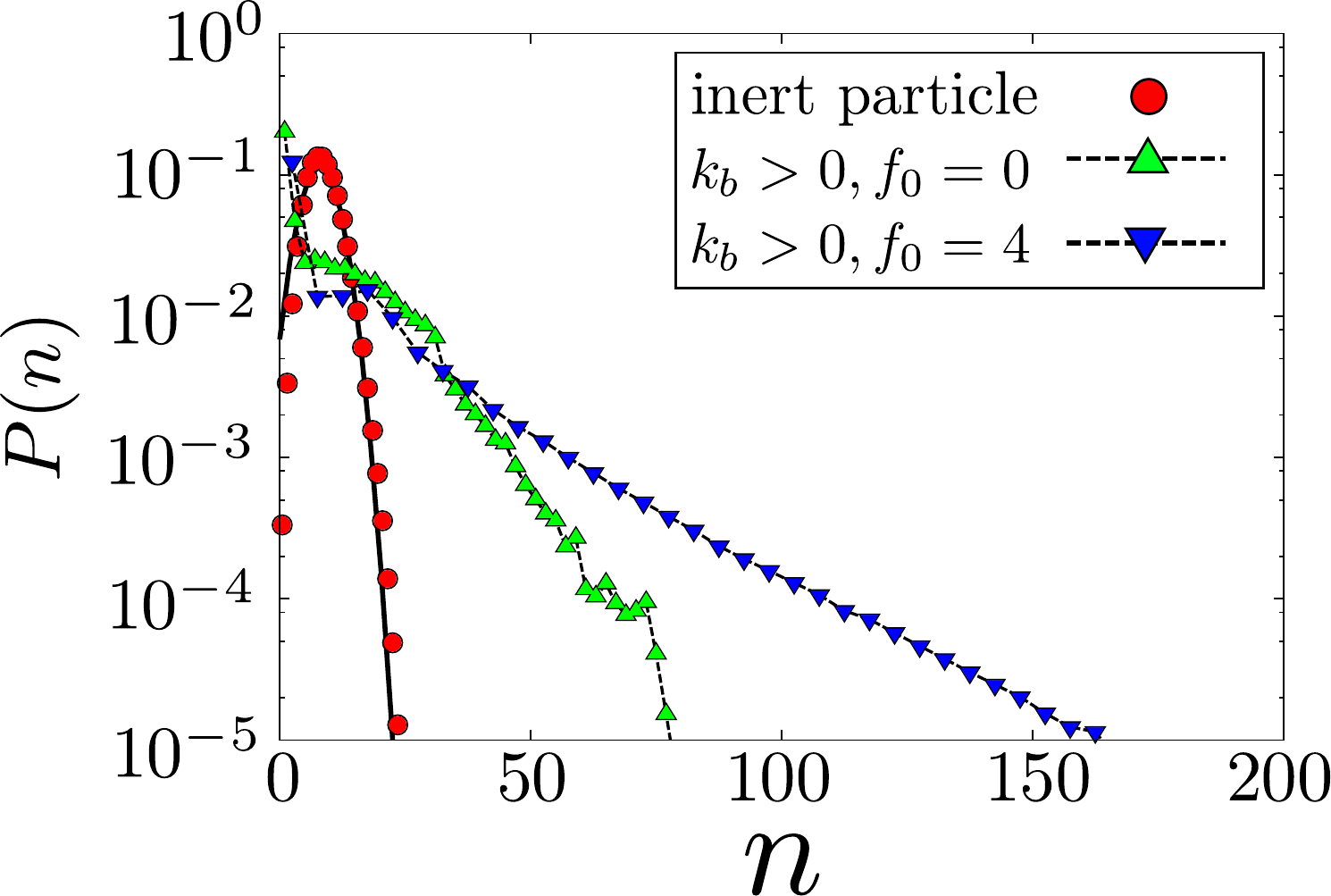}
\caption{Probability distribution of number density $P(n)$  for inert particles ($k_b=0$ : red $\bullet$), passive particles which can bind to the filaments at equilibrium ($k_b>0$, $f_0=0$ : green $\bigtriangleup$) and passive particles which can bind to the self-propelled filaments ($k_b>0$, $f_0=4$ : blue $\bigtriangledown$). For inert particles, $P(n)$ fits the virial expression Eq.\,\ref{numden} (dark line) with 
a second virial coefficient, $B_2=-0.99$. $P(n)$ picks up an exponential tail for particles that bind and unbind onto the filaments, that
moves towards the typical value as the active propulsion force gets larger.
}
\label{num_dist}
\end{figure}

The inert particles, in the dilute limit (Fig.\,\ref{num_dist}), has a probability distribution that resembles a
gas at temperature $T$ and the average number of particle in the blocks $\bar n$, namely, 
\begin{equation}
P(n)=\frac{1}{\sqrt{2 \pi \sigma^2}} \exp{\left[ -\frac{(n- {\bar n})^2}{2 \sigma^2}\right]}
\label{numden}
\end{equation} 
where $\bar n=\Omega \rho$. We have taken the variance of the above distribution to have a virial form, $\sigma^2 = \bar n - \frac{2 B_2}{\Omega} {\bar n}^2 + \ldots$ - a fit to the numerical data
gives $B_2=-0.99$ (dark line in Fig.\,\ref{num_dist}).

On the other hand, the probability distribution for passive particles picks up an exponential tail arising from the binding-unbinding statistics of the particles.
This exponential tail gets more pronounced when the filaments are made active, and 
which moves towards the typical value as the active propulsion force gets larger (Fig.\,\ref{num_dist}). 
This reflects the fact that for high driving, the typical particle is clustered. Both these results are entirely consistent with 
the experimental observations reported in Fig.\,4\,D of~\cite{darius}.

\subsection{Number fluctuations : crossover from anomalous to Brownian} 
\begin{figure}
\begin{center}
\includegraphics[height=0.70\linewidth]{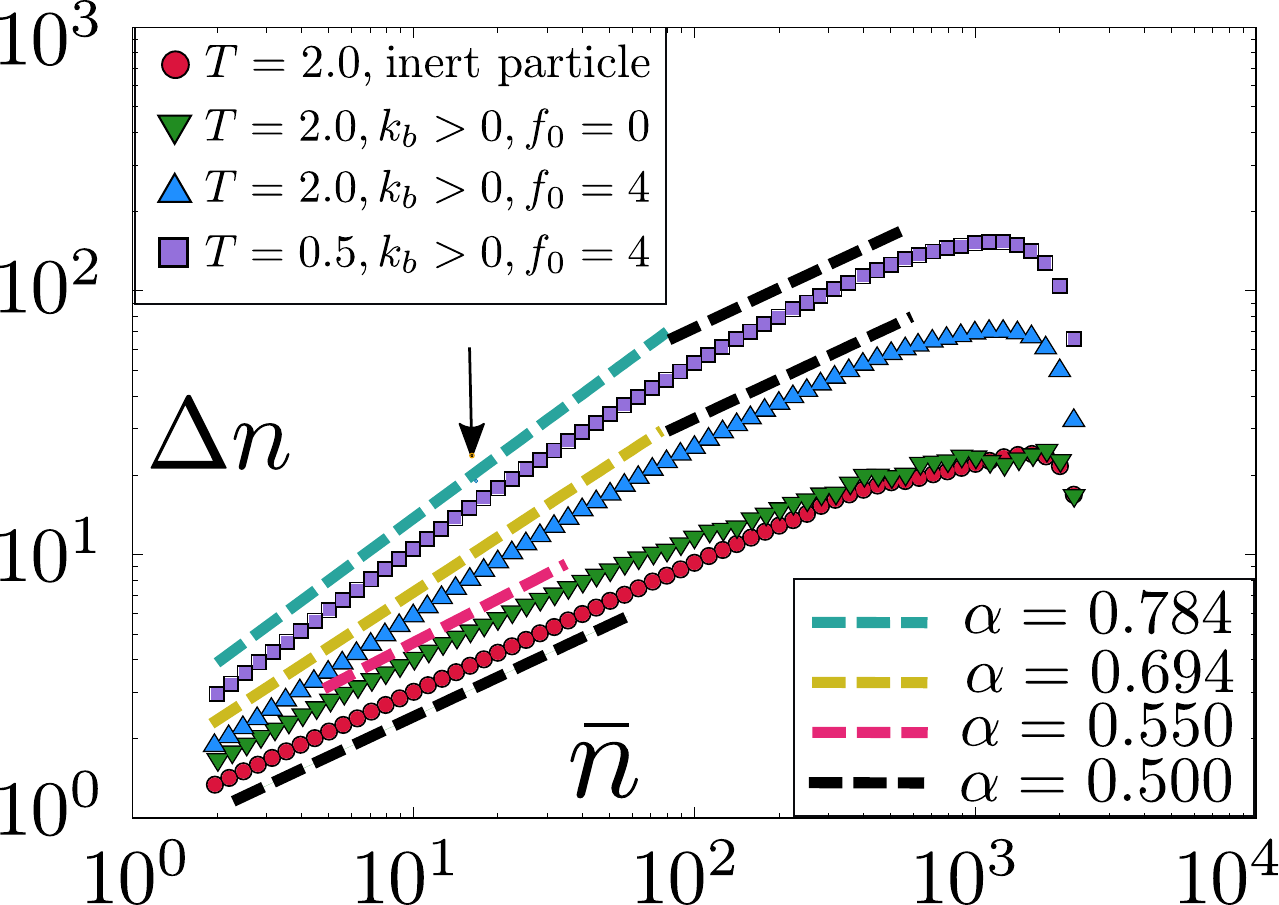}
\caption{Root mean square fluctuations of the number of passive particles $\Delta n$ versus the mean ${\bar n}$ for inert (red $\bullet$) and passive particles 
at two different temperatures,
$T=0.5$ ($\bigtriangleup$) and $T=2.0$ ($\square$).
Dashed lines indicate the local slope $\alpha$ in this log-log plot of $\Delta n \propto {\bar n}^{\alpha}$. The values of $\alpha$ indicated in the legend, show that while inert particles exhibit
normal fluctuations ($\alpha=0.5$), passive particles show large fluctuations at small ${\bar n}$ (with $\alpha$ depending on $T$ and $f_0$) that crosses over to 
normal fluctuations beyond a scale corresponding to the orientational correlation length (indicated by the arrow).\\
}
\label{giant}
\end{center}
\end{figure}

Note that the active system of filaments is in the isotropic phase and we should not expect to see giant number fluctuations normally associated with
active systems with global orientational order~\cite{toner-tu,simha,vijay}. However when we compute the root mean square fluctuations $\bigtriangleup n$ and mean 
${\bar n}$ of the number of passive particles over regions of ever increasing area, and plot them with respect to each other, we find that initially $\bigtriangleup n \propto {\bar n}^\alpha$
with $\alpha = 0.784$.
 Subsequently, as ${\bar n}$ increases, the variance scaling shows a cross over to $\alpha=0.5$.
This crossover occurs over a scale corresponding to the orientational correlation length, which can in principle be large, especially close to the
isotropic-nematic transition or high $f_0$. 
This is especially apparent in the high particle density regime, see Fig.\,\ref{giant} for particle density $\rho=0.05$ and filament density $c=0.02$.
This slow crossover is
 consistent with the experiments on the {\it in vitro} reconstitutions of actomyosin on a supported bilayer~\cite{darius}.

\section{Transport of passive advective scalars in an active medium}

We now study the transport of passive particles moving in the active medium. Because the filaments are orientationally disordered, the long time dynamics of the particles
is always diffusive. However the diffusion characteristics can change depending on the statistics of (un)binding to the active filaments.

\subsection{Typical trajectories}
\begin{figure}
\includegraphics[height=0.3\linewidth]{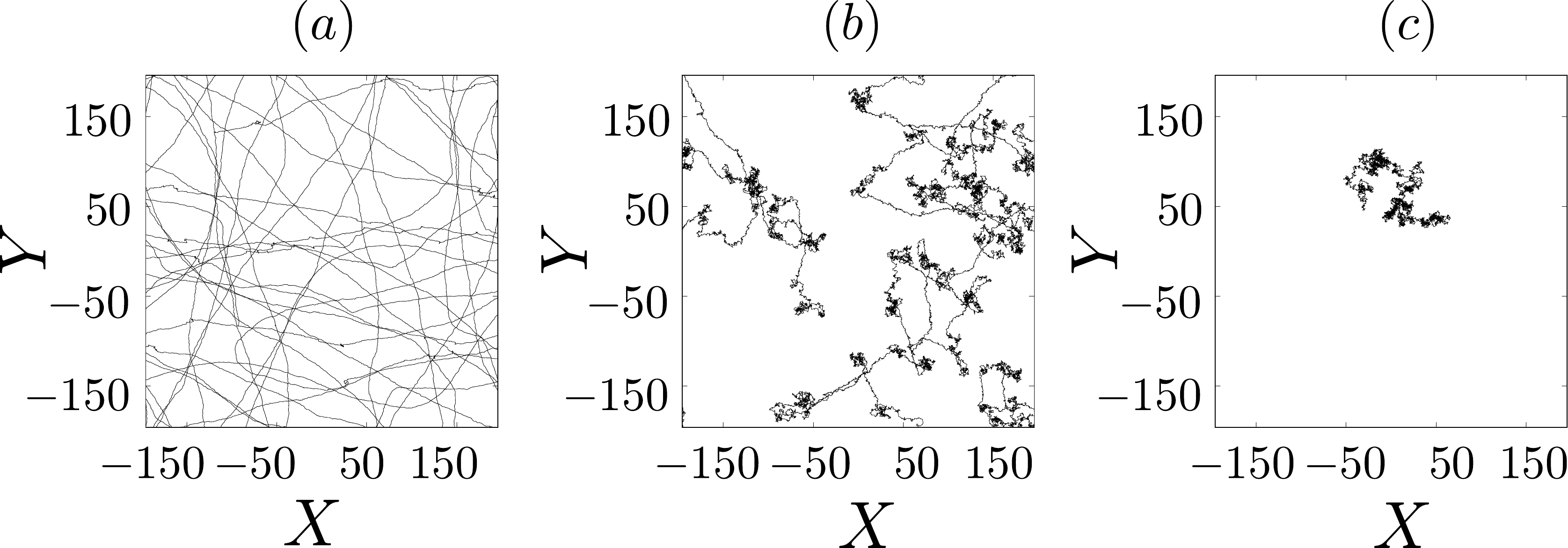}
\hspace{1mm}
\caption{Typical trajectories of passive particles for a fixed propulsion force $f_0=4.0$ at different temperatures - (a) $T=0.5$ (low : $k_BT/{\overline U} = 0.04$);
(b) $T=4.0$ (intermediate : $k_BT/{\overline U} = 0.32$) and  
(c) $T=10.0$ (high : $k_BT/{\overline U} = 0.8$).}
\label{trajec}
\end{figure}

The space-time trajectories of the passive particles show
three qualitatively different behaviours. At very low temperatures compared to ${\overline U} = K_sr_0^2/4$, 
a passive particle once bound
to a filament, rarely unbinds, and hence gets advected with the self-propelled filament (Fig.\,\ref{trajec}(a)).

The direction of 
advection changes because of thermal fluctuations and collisions between filaments. 
Increasing the temperature increases the probability of unbinding from the filament, whereupon the particle undergoes unrestricted thermal diffusion before 
binding again (Fig.\,\ref{trajec}(b)). At even higher temperatures, $k_BT/{\overline U} \approx  1$  the particles do not bind to the filaments and the motion is simple thermal diffusion
(Fig.\,\ref{trajec}(c)).

\subsection{Statistics of displacements and diffusion coefficient}

\noindent
{\bf Propensity distribution}.
The distribution of displacements  $\Delta x$ (along the ${\hat \x}$ direction) evaluated over a time window $t_w$ is called the
{\it propensity distribution}. This will depend on the statistics of binding/unbinding, which
in turn depends on the temperature $T$ and $f_0$, densities of filaments and particles, and of course on the time window $t_w$, which we fix at $t_w=10$. This can be obtained both from our Brownian dynamics 
simulation and, in the dilute limit, analytically.

In the dilute limit, one can obtain the form of this probability distribution from the stationary process describing  the particle vector-displacements in a small time interval $t_w$,
	\be
	{\bf r}(t_w)  =  {\bf r}(0) + \int_0^{t_w} \V(t') \,dt' 
	\label{position0}
	\ee
	where  $\V$ is the velocity of the tagged particle at time $t$, given by,
		\be
\V(t)  =  \phi(t) \, \frac{f_0}{\gamma}\, {\bf {\hat n}}(t) + (1-\phi(t)) \,{\boldsymbol \xi}(t),
\label{velocity}
\ee
where $\bf{\hat{n}}$ is the polar vector representing the orientation of the filament to which the particle is bound at time
$t$, ${\boldsymbol{\xi} }$ is the thermal noise, and 
$\phi(t)$ is the  telegraphic noise whose statistics is described in Sect.\,II\,D. The distribution of the particle displacements $\Delta x_{t_w}\equiv \left ({\bf r}(t_w) -  {\bf r}(0) \right)\cdot {\hat \x}$,
can be obtained by evaluating,
\be
P(\Delta x) = \langle  \delta\left(\Delta x-\Delta x_{t_w} \right) \rangle
\ee
where $\Delta x_{t_w}$ is obtained from Eq.\,\ref{position0} and the angular bracket denotes an average over the joint distribution of $\boldsymbol{\xi}$ and $\phi$. This can be evaluated by standard
techniques of Fourier transformation and cumulant expansion~\cite{kampen},
\be
\ln {\tilde P}(k) = \sum_{m=1}^{\infty} \frac{(i k)^m}{m !} \langle \left(\Delta x_{t_w}\right)^m  \rangle_c
\ee
where  ${\tilde P}(k)$ is the Fourier transform of $P(\Delta x)$. The $m$-th cumulants $\langle  \left(\Delta x_{t_w}\right)^m  \rangle_c$ can be evaluated from Eqs.\,\ref{position0},\,\ref{velocity}, knowing 
 that $\phi$ and ${\boldsymbol{\xi }}$ are independent stochastic processes. The distribution $P(\Delta x)$ is then obtained by taking the inverse Fourier transform of  ${\tilde P}(k)$.

However, in practice, the inverse Fourier transform of ${\tilde P}(k)$, for the stationary process Eq.\,\ref{position0}, has to be evaluated numerically.
Rather than do this, we 
 provide an alternate argument which gives more insight.

\begin{figure}
\includegraphics[height=0.36\linewidth]{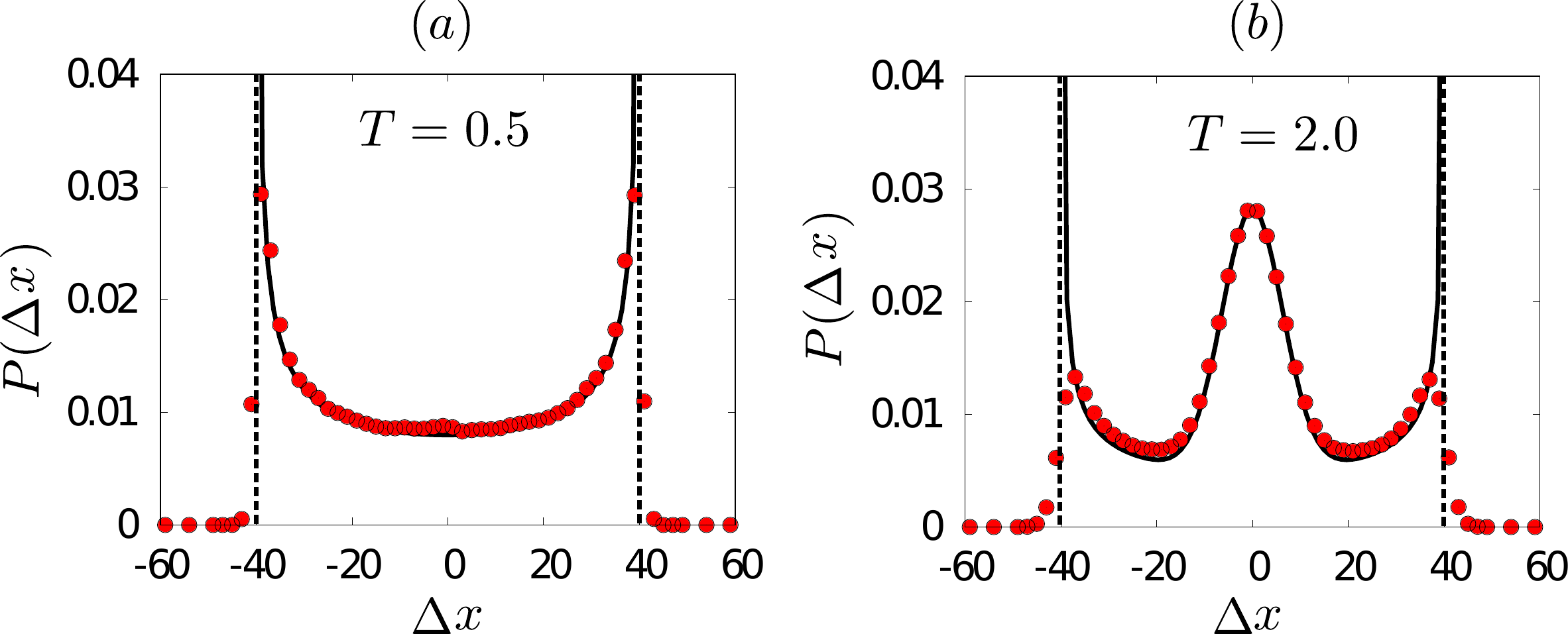}
\caption{Probability distribution $P(\Delta x)$ of the  displacements of passive particles, evaluated for fixed time interval 
$\bigtriangleup t =10$, at a propulsion force $f_0=4.0$ and temperatures (a) $T=0.5$ and (b) $T=2.0$. The central peak comes from the fraction undergoing thermal diffusion, while 
the side peaks come from the bound fraction undergoing active motion. 
The black filled line shows parameter-free fits to the approximate analytical form (Eq.\,\ref{propensity}), where we have estimated the value of $\langle \phi \rangle=0.9995$ at $T=0.5$ (mainly bound) and 
$\langle \phi \rangle=0.632$ at $T=2.0$.}
\label{propensity_dist}
\end{figure}

At low enough $T$, the 
passive particles are completely bound to the self-propelled filaments, and so as long as $t_w < \tau$, the 
orientational
correlation time of the filaments,
the particles get displaced by $\Delta{\bf{r}}=\frac{f_0 t_w}{\gamma_p} \bf{\hat{n}}$, where 
$\gamma_p$ is the friction coefficient and $\bf{\hat{n}}$ is the unit vector representing the average orientation of a filament during time interval
$t_w$. 
Since the filament orientation is uniformly distributed, the contribution to the step-size distribution from this process is 
$P(\Delta x)=\frac{\gamma_p}{\pi f_0 t_w} \frac{1}{\sqrt{1-(\frac{\gamma_p}{f_0}
\frac{\Delta x}{t_w})^2}}$. On the other hand, at very high $T$, the particles are completely unbound and undergo thermal diffusion,
for which the step-size distribution is $P({\Delta x})=\frac{1}{\sqrt{2 \pi \sigma^2}} \exp{[-\frac{
(\Delta x)^2}{2 \sigma^2}]}$, where $\sigma^2=\frac{2  k_B T}{\gamma_p} t_w$. 
We propose that at an intermediate $T$, the propensity distribution can be written as a  linear combination, weighted  by 
 the duty ratio $\langle \phi \rangle$, i.e., 
\bea
P(\Delta x)= \frac{\gamma_p}{\pi f_0 t_w} \frac{\langle \phi \rangle}{\sqrt{1-\left(\frac{\gamma_p}{f_0}\frac{\Delta x}{t_w}\right)^2}} + \nonumber \\
 \frac{1-\langle \phi \rangle}{\sqrt{2 \pi \sigma^2}} \exp{\left[-\frac{(\Delta x)^2}{2 \sigma^2}\right]}.
\label{propensity}
\eea
The agreement of this approximate analytical form, with no undetermined parameters, with the results of the Brownian simulation is quite reasonable, see Fig.\,\ref{propensity_dist}.
\\

\begin{figure}
\includegraphics[height=0.4\linewidth]{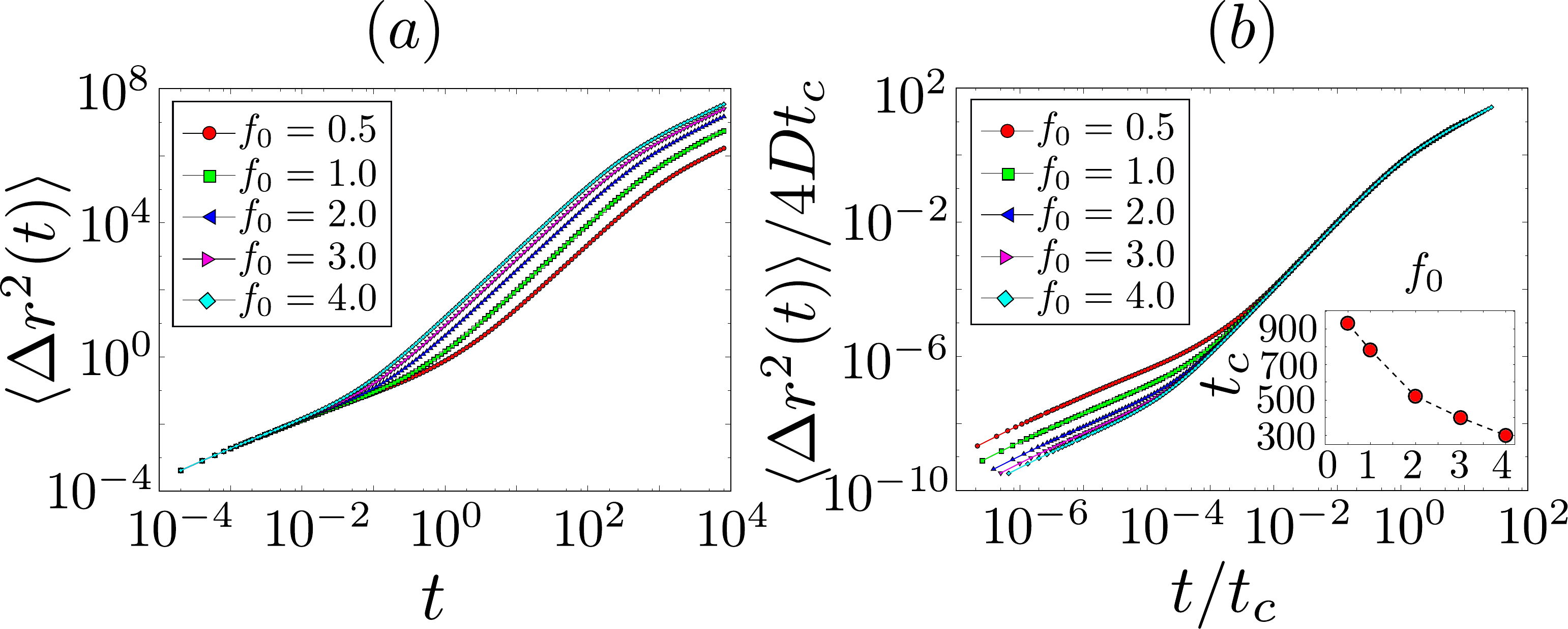}
\caption{(a) MSD of passive particles as a function of time at $T=0.5$ for different values of
self-propulsion force $(f_0)$ of the filaments. (b) Collapse of MSD using crossover time ($t_c$) and long time diffusion constant
for the same case. Inset shows crossover time $(t_c)$ as a function of self-propulsion force $(f_0)$ of the active filaments  for 
the same case.}
\label{MSD}
\end{figure}

\noindent
{\bf Mean square displacement}. From the statistics of the displacement we can compute the 
 mean square displacement (MSD) as $\langle \Delta {\bf r}^2(t) \rangle=\langle \frac{1}{N_p} \sum_i |{\bf r}_i(t_0+t) -{\bf  r}_i(t_0)|^2  \rangle$, where ${\bf r_i}$ is the position of $i$-th particle. This shows a change from a short time diffusive regime crossing over to a long time diffusive regime via an
 intermediate super-diffusive regime (Fig.\,\ref{MSD}(a)). We estimate the second crossover time $t_c(T,f_0)$ from the super-diffusive to late time diffusion $D$, by fitting the simulation data to $\langle \Delta {\bf r}^2(t) \rangle= 4Dt\,[1-\exp(-t/t_{c})]$~\cite{libchaber},
using which
we can collapse the MSD data for different values of active propulsion $f_0$ (Fig.\,\ref{MSD}(b)). From this we see that $t_c$ decreases with
$f_0$ (Fig.\,\ref{MSD}(b) inset). This is because the filament orientation decorrelates on account of collisions, whose frequency increases with $f_0$.  In experimental systems  where  hydrodynamics plays a crucial 
role~\cite{libchaber}, this dependence of $t_c$ on
$f_0$ may be different.
In situations where the crossover $t_c$ is large, the apparent super-diffusion behaviour would last for many decades in time.
We can then fit the MSD to  $\langle \Delta {\bf r}^2(t) \rangle \sim t^\beta$ to obtain an super-diffusion exponent $\beta > 1$ - 
we find that $\beta=1.95$ at $T=0.5$ and $\beta=1.77$ at $T=2.0$, for a propulsion force $f_0=4.0$.\\

\begin{figure}
\begin{center}
\includegraphics[height=0.39\linewidth]{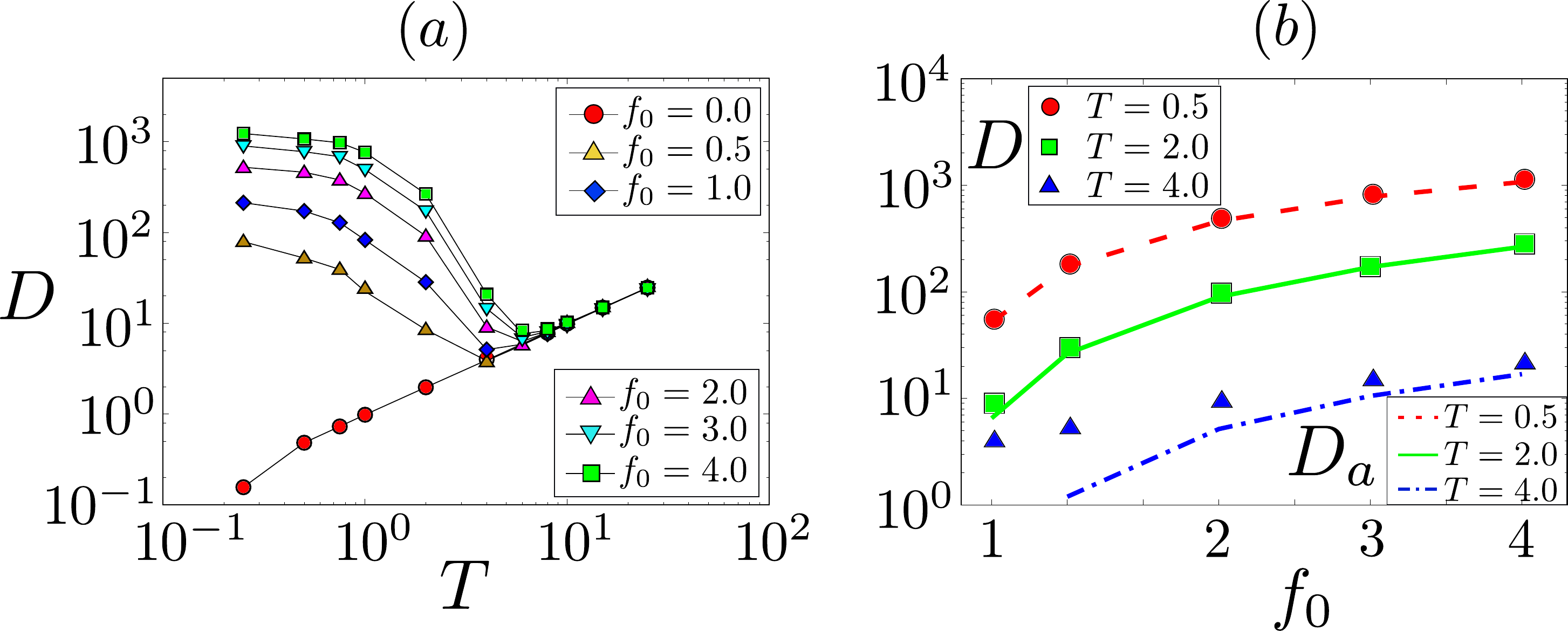}
\caption{(a) Diffusion coefficient ($D$) of tagged passive particles has been plotted as a function of $T$ for different values of
 $f_{0}$ of the filaments. At low temperature regime $D$ is weakly dependent on $T$ (signature of {\it `active diffusion'}).
At intermediate regime $D$ decreases then again increases at high temperature regime where thermal diffusion dominates.
(b) We have plotted total diffusion coefficient $D$ (with points) and {\it active diffusion coefficient} $D_a$ ( with line) with $f_0$ for different $T$.  In low temperature and high activity 
regime the difference between $D$ and $D_a$ is insignificant, as we increase $T$ and decrease $f_0$ the difference becomes prominent.}
\label{diffusion}
\end{center}
\end{figure}

\noindent
{\bf Temperature and activity dependence of MSD}.
We now compute the late time diffusion coefficient of the tagged particles, $D = \lim_{t \to \infty}\langle \Delta {\bf r}^2(t) \rangle/4t$, for different $T$ and $f_{0}$. 
For a fixed $f_0$, one might expect that at low temperatures $D$ is weakly dependent on (or even independent of)
temperature because a particle once bound to the filament remains so and undergoes {\it active diffusion} as it is transported by the filament
(Fig.\,\ref {diffusion}). 
As we increase the temperature, $D$ decreases, since a particle spends less time, on an average, bound to the filament (recall we have set $\gamma_b=\gamma_p$).
At high temperatures, the particles are predominantly unbound, and hence  $D$ resembles that of an inert particle, which increases linearly with temperature.
This is indeed what we see from a direct numerical simulation of the Brownian dynamics trajectories of a tagged particle (Fig.\,\ref{diffusion}).

From the stationary process, Eq.\,(\ref{position0}), the MSD of the tagged passive particle,
\be
\la \delta {\bf r}^2(t) \ra  =  \, \int_0^t \int_0^t  \la \V(t') \cdot \V(t'')\ra dt' dt''  \, ,
\label{msd0}
\ee
 immediately gives the diffusion coefficient, 
\be
D = \frac{1}{2} \int^{\infty}_{0} \la \V(t) \cdot \V(0)\ra dt.
\label{diff0}
\ee

\begin{figure}
\begin{center}
\includegraphics[height=0.88\linewidth]{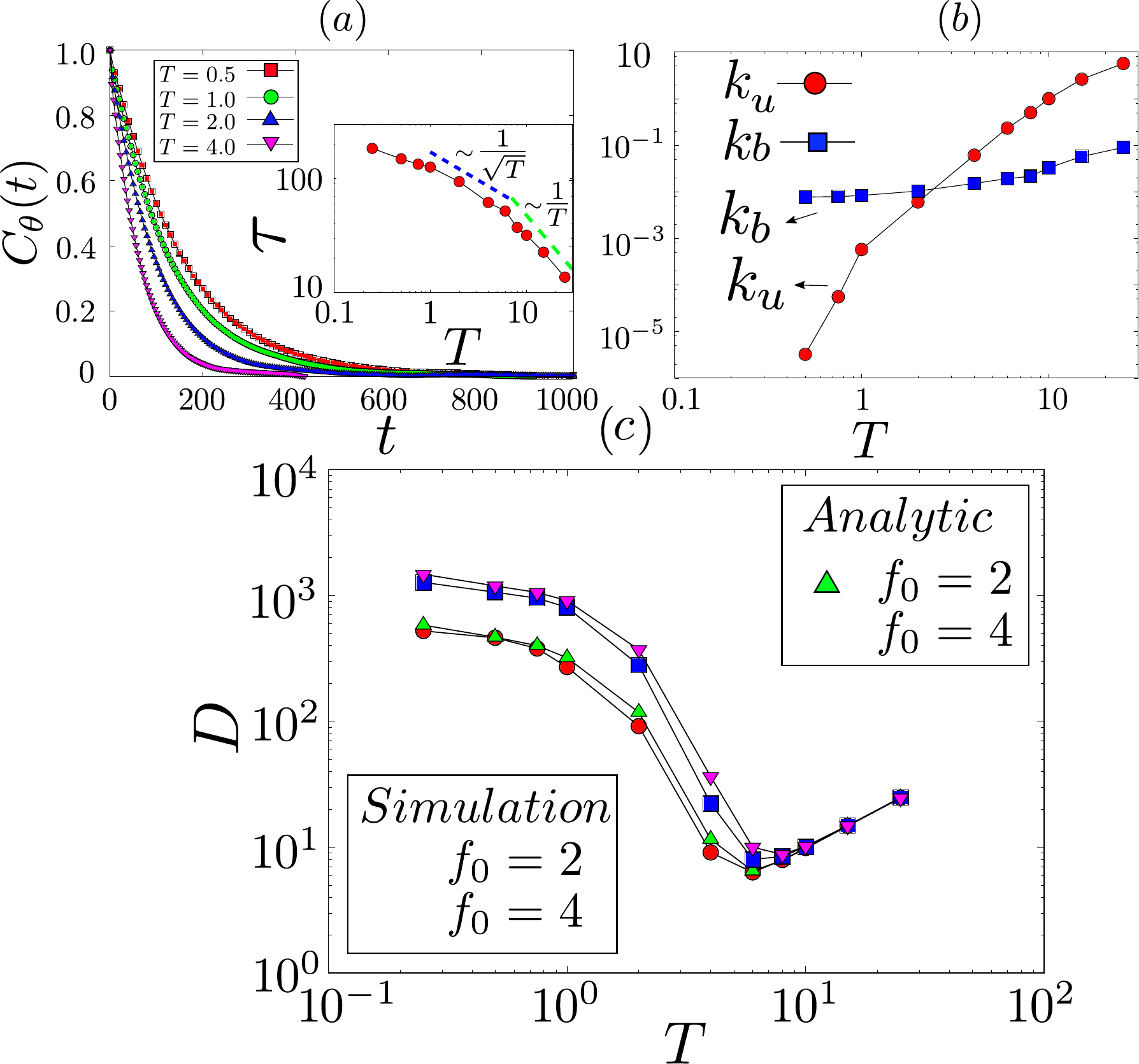}
\caption{(a) Temporal correlation of the direction of persistent movement of the filaments, measured by the velocity orientation correlation function $C_\theta(t)$, at different temperatures, with
the filament density $c=0.01$ and propulsion force $f_0=4$.  Inset shows the corresponding orientation correlation time $\tau$ as a function of $T$. (b) Temperature dependence 
of binding ($k_b$) and unbinding ($k_u$) rates calculated from the statistics of the telegraphic noise $\phi(t)$. (c) Both, simulation and analytic calculation of diffusion coefficient ($D$) of 
tagged passive particles has been plotted as a function of temperature ($T$) for two different activities ($f_0=2,4$) for filament density $c=0.01$. \\
}
\label{analytic}
\end{center}
\end{figure}

Using Eq.\,(\ref{velocity}), we see that the diffusion coefficient of the bound particle is given 
by the correlations of $\V_a$, which  is given by (Fig.\,\ref{analytic}(a)),
 \bea
   \la \V_a(t) \cdot \V_a(0)\ra &  =  &  \frac{f_0^2}{\gamma_p} \la \cos(\theta(t)-\theta(0)) \ra \nonumber \\
   & =  & \frac{f_0^2}{\gamma_p} e^{-\,\mid t\mid/\tau}
   \label{correl}
 \eea
  The diffusion coefficient can now be simply evaluated,
  \bea
  D=\frac{f_0^2 \tau k_b(\tau k_b +1)}{2 \gamma_p^2(k_u+k_b)[(k_u+k_b)\tau+1]}+\frac{k_B T k_u}{\gamma_p(k_u+k_b)}
  \label{diff}
  \eea

  To plot $D$ versus $T$ and $f_0$, we need to know the values of $k_b$, $k_u$ (equivalently $\langle \phi\rangle$, $t_{sw}$) and $\tau$, which depend on the temperature and density, and which
  we obtain from our simulations.
  We then compare this semi-analytical form to the direct numerical computation of the diffusion coefficient from the Brownian dynamics trajectories (Fig.\,\ref{analytic}).
  The agreement between the two is excellent. These observations are entirely consistent with the results of~\cite{saha}.

  It might be objected that in our analysis we have treated the active propulsion as an independently tunable parameter, thus precluding the possibility that the 
  activity itself may be temperature dependent. However, as we saw in~\cite{Goswami}, and as noted elsewhere~\cite{spudich,levy}, the actomyosin contractile processes taken as a whole, 
  appear to be independent of temperature in the physiological range,
$24^{\circ}-37^{\circ}$C.

\begin{figure}
\includegraphics[height=0.88\linewidth]{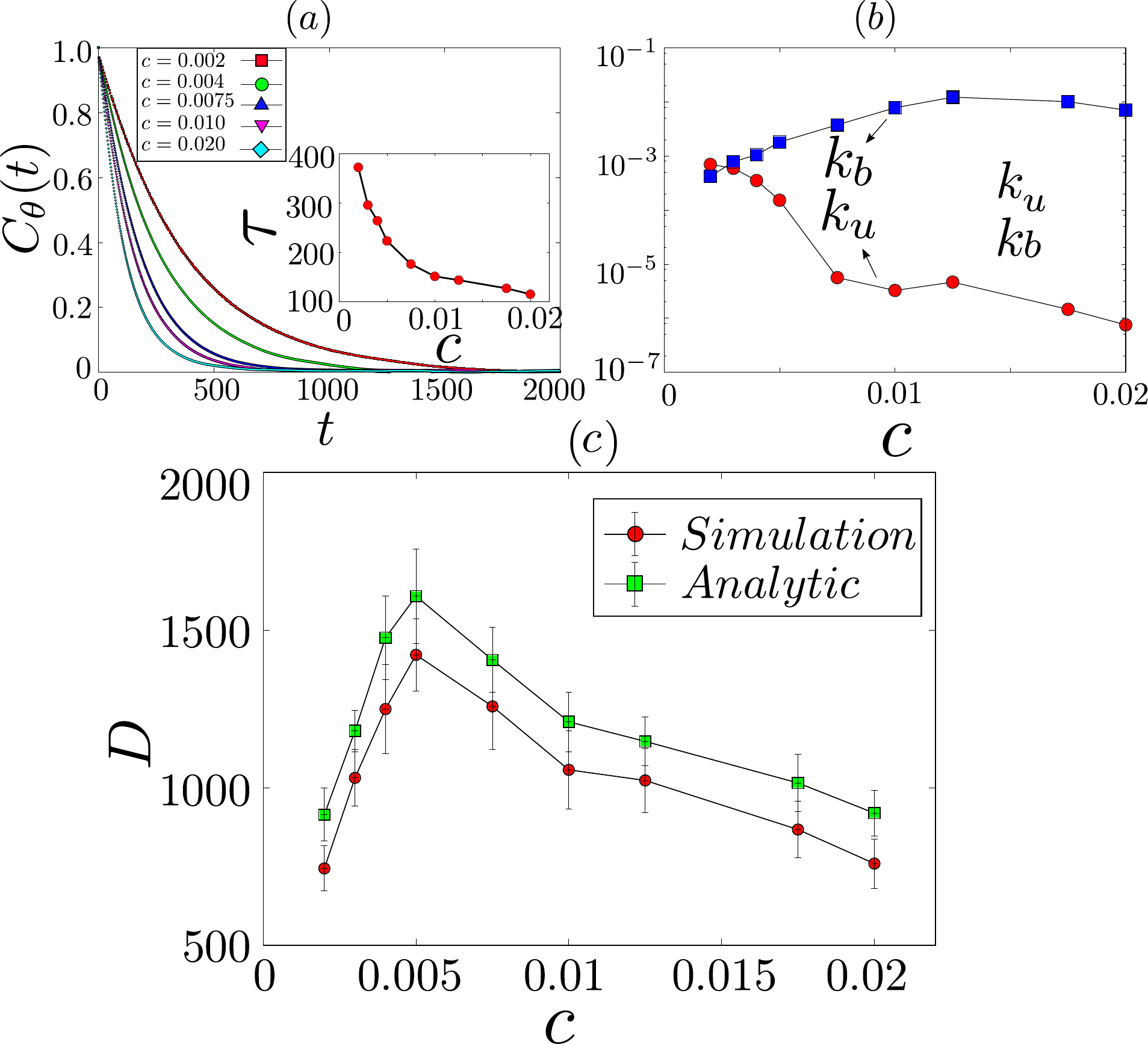}
\caption{(a) Temporal correlation of the direction of persistent movement of the filaments, measured by the velocity orientation correlation function $C_\theta(t)$, 
with different filament density ($c$), at temperature $T=0.5$ and propulsion force $f_0=4$. Inset shows the corresponding orientation correlation time $\tau$ as a function of $c$. (b) Dependence 
of binding ($k_b$) and unbinding ($k_u$) rates on filament density $c$, calculated form the statistics of the telegraphic noise $\phi(t)$. (c) Both, simulation and analytic calculation of diffusion coefficient ($D$) of 
tagged passive particles has been plotted as a function of filament density ($c$) at $T=0.5$ and for $f_0=4$.\\
}
\label{optimal}
\end{figure}

Figure\,\ref{optimal} shows the dependence of $D$ on filament concentration $c$, both from direct simulations and from the analytical form using the values of 
$k_b$, $k_u$ (equivalently $\langle \phi\rangle$, $t_{sw}$) and $\tau$,
from simulations. This shows optimal transport at a specific filament concentration; the orientational decorrelation time is smaller at higher filament concentration, due to higher collision
frequency.

\section {Viscosity stratification and its effect on membrane diffusion}
So far, our study of transport of passive molecules in an active medium has been restricted to two dimensions. However as we discussed in Sect.\,I, the cell surface is a composite of a bilayer membrane and a 
thin actomyosin cortex. Thus while the proteins move on the cell membrane, the actively driven actin moves in the actomyosin cortex. The viscosities of these two layers are significantly different, with the
bilayer membrane having a viscosity which is an order of magnitude larger than the cortex ($\approx0.86$\,Pa\,s~\cite{kamm}).
Indeed the local viscosity of a multicomponent membrane can be quite heterogeneous - for instance the particle mobility within the so-called ``membrane rafts'' or liquid-ordered regions on the cell membrane
can be very different from those within liquid-disordered regions. Moreover the local cortical viscosity depends on local actin, myosin and cross-linker concentrations.
 How does this viscosity mismatch affect the actively driven transport of passive molecules?

\begin{figure}[H]
\includegraphics[height=0.35\linewidth]{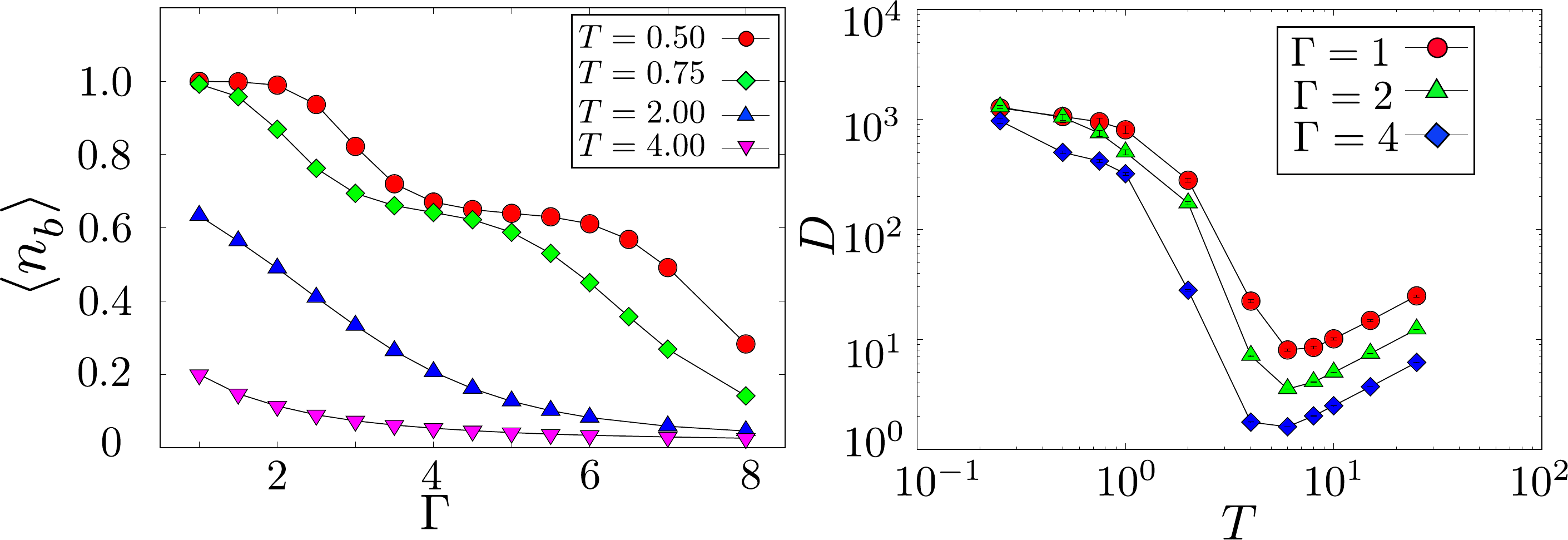}
\caption{(a) Fraction of bound particles $\langle n_b \rangle$ decreases with $\Gamma=\gamma_p/\gamma_b$ over a range of temperatures. 
(b) Diffusion coefficient versus temperature at  different values of viscosity mismatch parameter $\Gamma$. In both figures, the propulsion force has been 
fixed at $f_0=4.0$.}
\label{mismatch}
\end{figure}

To address this important issue within our simulation, we vary the ratio of the friction coefficients $\Gamma = \gamma_p/\gamma_b$ in Eqs.\,\ref{particle},\,\ref{bead}. 
We find that the mean fraction of passive particles bound to filaments $\langle n_b \rangle$
 decreases with increasing $\Gamma$ over a range of $T$ and $f_0=4.0$ (Fig.\,\ref {mismatch}(a)).
This is an interesting observation, since one might have naively thought that $\langle n_b \rangle$ is solely governed by binding-unbinding, a purely equilibrium process and hence 
independent of relative viscosities. However, we see that the drag induced by the imposed viscosity stratification (a nonequilibrium feature), can ``peel-off'' particles from the filaments.
It is not clear to us why we see a shoulder at intermediate values of $\Gamma$ for low enough temperatures (Fig.\,\ref {mismatch}(a)).

This is reflected in changes that we observe in the measured diffusion coefficient $D$, as it decreases with increasing $\Gamma$ at different $T$ (Fig.\,\ref{mismatch}(b)).
As can be seen, the active-diffusion regime at low temperatures becomes significantly more temperature dependent as the viscosity mismatch $\Gamma$ increases.

The results of this section are not purely academic, on the contrary taken together they pose an interesting possibility that by tuning local viscosity mismatch, 
for instance by locally recruiting the so-called ``membrane rafts'' or liquid-ordered regions on the cell membrane or by  locally regulating the concentrations of actin, myosin or cross-linkers,
the living cell surface could control  the clustering and transport of
specific membrane proteins.

\section{Discussion}
We had earlier shown that a coarse grained active hydrodynamics description of the active composite cell surface, successfully explains the statistics of clustering 
of membrane proteins capable of binding to the cortical actomyosin in living cells~\cite{raomayor,kripa}. Such a description make predictions regarding the statistics of density fluctuations and
transport of such actin-binding membrane proteins, which were verified in experiments~\cite{kripa,saha}. Following this we were able to recapitulate much of this behaviour in a minimal
{\it in vitro} system comprising  a thin layer of short actin filaments and Myosin-II minifilaments on a supported bilayer~\cite{darius}. The success of this approach has motivated us to do a
agent-based Brownian dynamics simulation using these minimal ingredients - that of a collection of passive molecules which bind/unbind to actin filaments and move in this active medium in two dimensions.

The  results obtained  here, based on simulations and analytical calculations,
 are in qualitative agreement with 
the experiments both {\it in vivo} and {\it in vitro}. For instance, the exponential tails appearing in the probability distribution of the number (Fig.\,\ref{num_dist}) and the 
scaling of the variance of the number (Fig.\,\ref{giant}) is precisely the behaviour seen in our earlier {\it in vitro} experiments.
In addition, we show how activity induced clustering of passive particles (Fig.\,\ref {gr_density}) arises naturally from such a minimal description.

We have also studied transport of passive particles moving in this active medium, and find that there is a crossover from an intermediate time super diffusive to late time
diffusive behaviour as a consequence of active driving (Fig.\,\ref{MSD}(a)). The transport behaviour shows a striking  dependence on temperature and active forcing - at low temperatures
the diffusion coefficient is insensitive to temperature, and crosses over to a linear temperature dependence at higher temperatures, in qualitative agreement with 
experiments~\cite{saha}.

Finally, recognising that the viscosity of the cortical layer is different from that of the membrane, we show that a friction coefficient mismatch has a strong effect on the mean
number of bound particles and the diffusion coefficient. This is a consequence of the drag induced by the imposed viscosity stratification, which results in a 
 ``peeling-off'' of the particles from the filaments. This opens up the possibility of  local tuning of viscosity mismatch, for instance by locally recruiting the so-called ``membrane rafts'' or
  liquid-ordered regions on the cell membrane or by locally regulating the concentrations of actin, myosin or cross-linkers. This could result in yet another mechanism by the cell surface might locally
 control the clustering and transport of specific membrane proteins. We hope that some of these predictions can be tested in future experiments.

\section{Acknowledgement}
We thank K. Husain, R. Morris and D. Banerjee for critical inputs. 
MR thanks S. Mayor and D. Koster for long years of collaboration.
SKRH and RM thank the Simons Centre at NCBS for computational facilities and hospitality.
RM acknowledges financial support from CSIR, India.

\end{document}